\documentclass[a4paper]{jpconf}
\usepackage{graphicx}
\usepackage{xspace}
\usepackage[noabbrev]{cleveref}
\usepackage{footnote}
\usepackage{tikz}
\newcommand{\RooFit}{\texttt{RooFit}\xspace}
\newcommand{\RooStats}{\texttt{RooStats}\xspace}
\newcommand{\HistFactory}{\texttt{HistFactory}\xspace}
\newcommand{\ROOT}{\texttt{ROOT}\xspace}
\newcommand{\ie}{\textit{i.e.}\xspace}

\begin{document}

\title{Making RooFit Ready for Run 3}
\author{S Hageb\"ock$^1$ and L Moneta$^1$}
\address{$^1$ CERN, 1211 Geneva 23, Switzerland}
\ead{stephan.hageboeck@cern.ch}

\begin{abstract}
\RooFit and \RooStats, the toolkits for statistical modelling in \ROOT, are used in most searches and measurements at the Large Hadron Collider. The data to be collected in Run 3 will enable measurements with higher precision and models with larger complexity, but also require faster data processing.

In this work, first results on modernising \RooFit's collections, restructuring data flow and vectorising likelihood fits in \RooFit will be discussed. These improvements will enable the LHC experiments to process larger datasets without having to compromise with respect to model complexity, as fitting times would increase significantly with the large datasets to be expected in Run 3.
\end{abstract}

\section{Introduction}
\begin{figure}[t]
\centering
\includegraphics[width=0.9\textwidth]{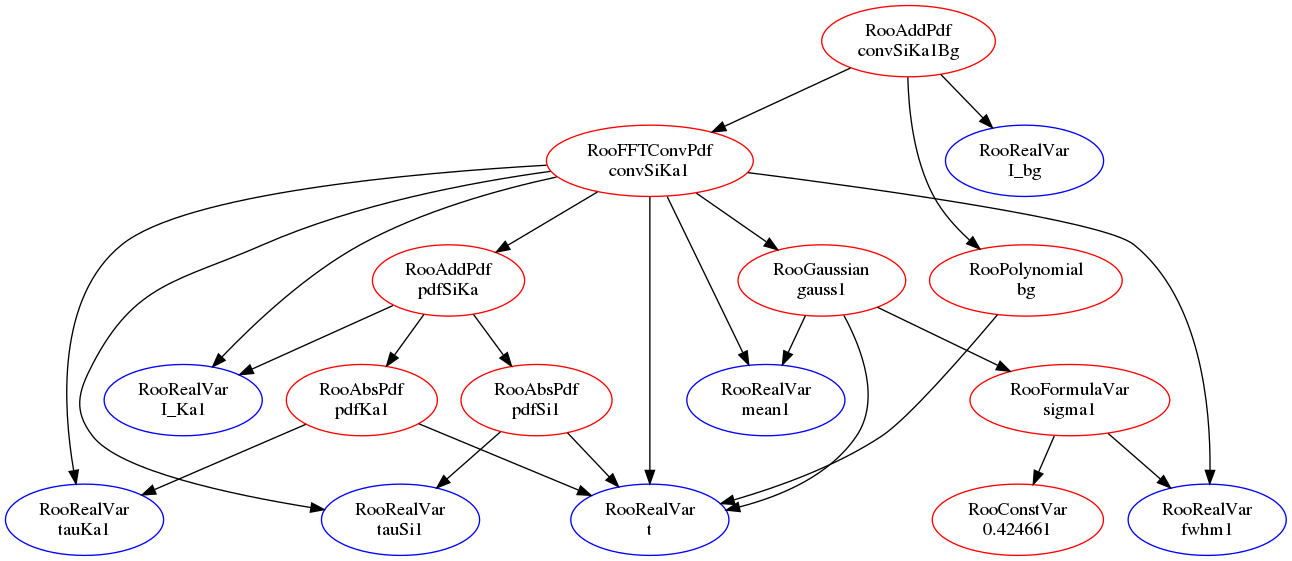}
\caption{\label{PDFTree}A \RooFit likelihood model\protect\footnotemark. This model represents the sum of a signal and background PDF,
where the former is a convolution of other PDFs, the latter is a polynomial distribution.
Likelihood models are implemented as tree-like structures of nodes that can be evaluated to real values.
Blue nodes represent parameters or observables, red nodes, which depend on the values of other nodes, represent functions or PDFs.
}
\end{figure}

\texttt{RooFit}~\cite{RooFit} is a C++ package for statistical modelling distributed with \texttt{ROOT}~\cite{ROOT}.
\RooFit allows to define computation graphs connecting observables, parameters, functions and PDFs in order to compute likelihoods and perform fits to data. An example of such a computation graph is shown in \cref{PDFTree}. Every node of the graph can be
evaluated to a real value (\textit{i.e.}, real-valued number), which for a PDF denotes the probability to find a data event with the given
values of observables, for a given model and its set of parameters.

\texttt{RooStats} is a collection of statistical tools to perform statistical tests with \texttt{RooFit}
models (\textit{e.g.} Toy Monte Carlo, setting limits). Further, \texttt{HistFactory} provides tools to create \RooFit models from a collection of \ROOT histograms.

\RooFit was originally developed for the BaBar collaboration, but later picked up by many others. Its central parts were designed for
single-core processors, and were neither optimised for large caches nor SIMD computations. This work stands at the beginning of efforts to
modernise \RooFit to speed up fits, and make it more accessible from both C++ and Python. The features to be described in the following two sections will be released in \texttt{\ROOT 6.18}.

\section{Modernising \RooFit's Internal Collections\label{sec:STLCollections}}
In \RooFit, computation graphs and sets/lists of functions, PDFs, observables and parameters are saved with the help of \texttt{RooAbsCollection},
the base class of \RooFit's main collections \texttt{RooArgSet} and \texttt{RooArgList}. Internally, these were using a
linked list with optional hash lookup.

The most common operation on these collections during fitting, \ie, repeated evaluation of the computation graph,
is iteration. Less frequent operations are appending and finding elements, and collections are very rarely sorted
or modified. This favours array-like data structures, and indeed 
the linked lists were identified as a bottleneck for fits.

\begin{figure}[h]
\vspace{-5mm}
\hspace{5mm}
\begin{minipage}[b]{0.5\textwidth}
\centering
\includegraphics[width=0.95\textwidth]{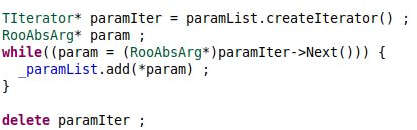}
\vspace{-2mm}
\caption{\label{itBefore}Iterating through a \RooFit collection with old interface.}
\end{minipage}\hfill%
\begin{minipage}[b]{0.4\textwidth}
\centering
\includegraphics[width=0.95\textwidth]{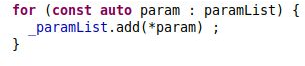}
\vspace{0.55mm}
\caption{\label{itAfter}Iterating through a \RooFit collection with new interface.}
\end{minipage}
\hspace{5mm}
\end{figure}

\begin{figure}[t]
\centering
\begin{tikzpicture}%
\node[anchor=south west, inner sep=0] (X) at (0,0){\includegraphics[trim=0 0 20mm 0,clip,width=0.9\textwidth]{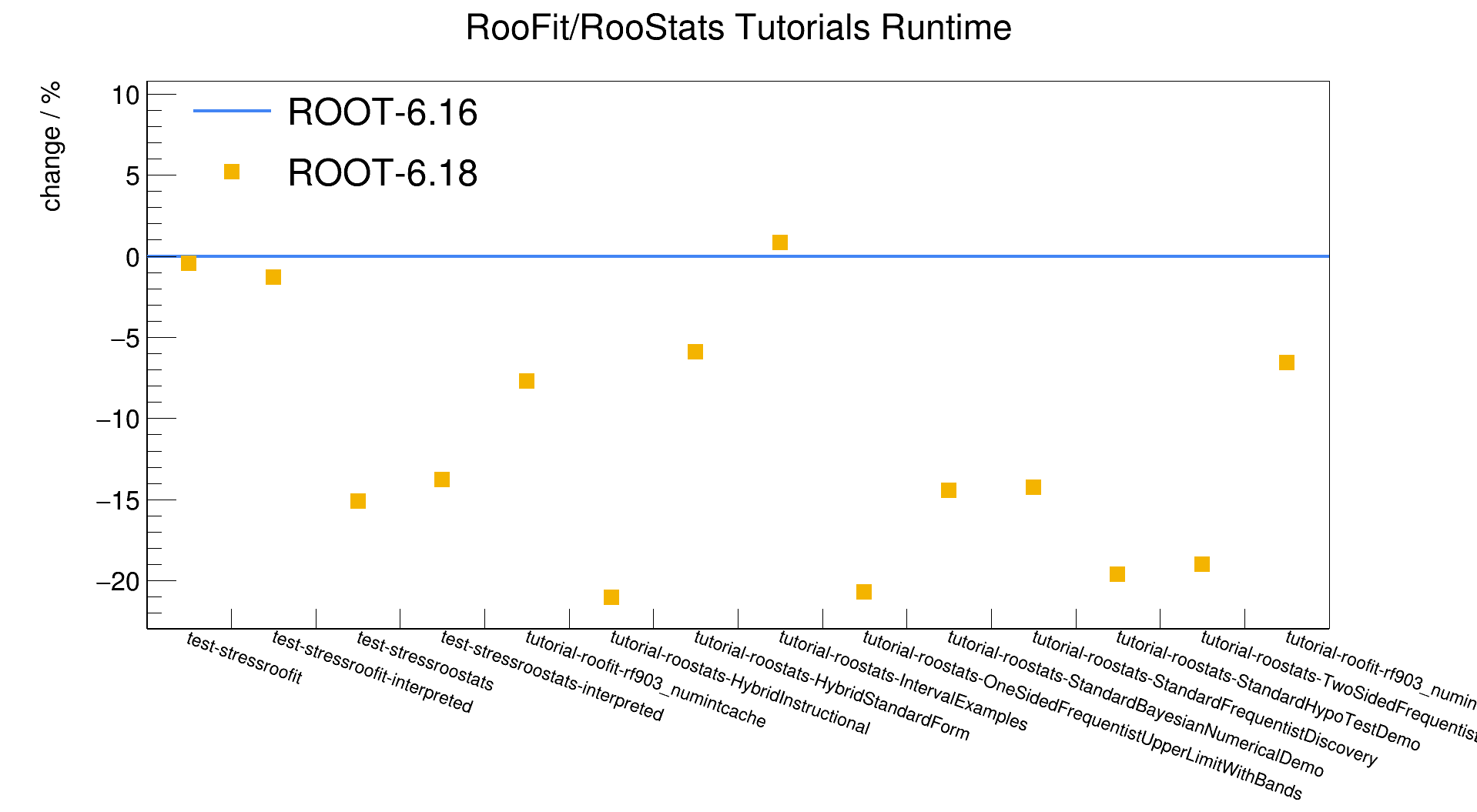}};%
\begin{scope}[x={(X.south east)},y={(X.north west)}]%
%% This makes all measurements with no units into fractions of the width
%% and height of the graphic contained in the node.
\node[anchor=north east,rotate=90,fill=white] (Z) at (0.015,0.92) {%
    \scriptsize Change in run time vs. \texttt{ROOT 6.16} / \%%
};
\end{scope}%
\end{tikzpicture}
\caption{\label{speedupSTL}%
Comparison of run times for typical \RooFit workflows between \ROOT \texttt{6.16} and \texttt{6.18}. Frequent use of STL-like iterators leads to a
speed up of $20\,\%$, heavy use of legacy iterators to a slow down.
}
\end{figure}

Therefore, the linked list in \texttt{RooAbsCollection} was replaced by a \texttt{std::vector}, mostly to speed up iterating. 
\texttt{RooAbsCollection} was further provided with an STL-like interface (\texttt{size, begin, end}) to enable range-based for loops.
This speeds up forward iteration by about $20\,\%$, and random access completes in constant time.
\footnotetext{This figure was obtained using the \texttt{graphVizTree()} export supported by all nodes in a \RooFit graph.}
The \cref{itBefore,itAfter} compare the old and new interface. The STL-like interface allows to
reduce heap allocations, and replaces while loops with non-local variables by range-based for loops. This reduces code clutter and the danger of memory
leaks, dangling pointers and variable shadowing. It will also facilitate
generating \ROOT's automatic Python bindings to iterate through \RooFit's collections in Python. \Cref{speedupSTL} compares
run times for typical \RooFit workflows\footnote{These are a selection of representative \RooFit tutorials.} between \texttt{\ROOT 6.16} and \texttt{6.18}.
Depending on how often collections are iterated, the speed up varies between $5$ and $21\,\%$. Tests with an ATLAS likelihood 
model~\cite{Hbb} yielded a speed up of $19\,\%$.

\subsection{Backward Compatibility}
The old interface of \texttt{RooAbsCollection} and its subclasses exposed three kinds of iterators to users, one of which
is shown in \cref{itBefore}. All three are being used in \RooFit, and it is likely that \RooFit's users also use all three.
Removing any of these would therefore break user code.

To provide backward compatibility, these legacy iterators were re-implemented to also work with the new collections. Since the old \RooFit
collections were based on a linked list, the legacy iterators need to remain valid also when reallocations happen.
Therefore, the legacy iterators hold a reference to the new collection, and use index access to
iterate through it. This makes them tolerant against reallocations, but inserting or deleting elements \textit{before} the current
position is not supported, unlike for linked lists. Not a single instance of such usage was found in \RooFit, though. Nevertheless, run-time checks were added
to warn users if they insert/delete before the current iterator. These are performed only if \ROOT is compiled with assertions enabled, and only for legacy iterators.

All legacy iterators further need to work both with the original \texttt{RooLinkedList} and
with the STL-based collections\footnote{The \texttt{RooLinkedList} will be deprecated, but not all instances of its usage have been removed.
It might further be used in user code.}. They were therefore implemented as adapters to a polymorphic iterator interface, which
supports both the \texttt{RooLinkedList} and the counting iterator. This means that all legacy iterators will work irrespective of
reallocations and the actual type of the collection, but they are slower than the fastest original \RooFit iterator, because they
require a heap allocation to polymorphically switch between different  backends. The slow down observed for one workflow in
\cref{speedupSTL} is caused by this.

Nevertheless, instances of slow legacy iterators are easily found using profiling tools, and can be replaced by STL
iterators by changing the code as shown in \cref{itBefore,itAfter}, which leads to a speed up of $20\,\%$. The most critical iterators in \RooFit have already been replaced,
and less critical iterators will be replaced as the modernisation of \RooFit continues.

\section{Faster \HistFactory Models\label{sec:HistFactory}}
\HistFactory~\cite{HistFactory} is a toolkit to create \RooFit models from a collection of histograms. It supports multiple channels,
multiple signal and background samples, sample scale factors, systematic uncertainties for shape and normalisation differences in histograms, and allows to implement combined measurements of parameters such as a signal strength.

To parametrise systematic uncertainties using histograms, users supply three histograms of the same distribution: the nominal distribution
and two histograms representing the $\pm 1\, \sigma$ uncertainty. Given that these have to be evaluated for multiple (sometimes hundreds) of
systematic uncertainties, for multiple samples and multiple channels, several thousands of histograms might have to be analysed.

When \HistFactory was implemented, move semantics or shared pointers were not available. The authors therefore resorted to copying
histograms, leading to a large overhead of copying and deleting histograms. Performance-critical sections of 
the \HistFactory code were therefore revisited, and move semantics as well as shared pointers implemented. This speeds up creating a likelihood
model for an ATLAS measurement~\cite{Hbb} by more than ten times, with identical results. This model comprises $10\,832$
histograms, $28$ channels and $253$ systematic uncertainties, and was constructed in $150\,\mathrm{s}$ instead of $1\,800\,\mathrm{s}$. Fits using this model furthermore
converged $20\,\%$ faster because of the optimisations discussed in \cref{sec:STLCollections}.

\section{Batched Likelihood Computations}
A bottleneck for likelihood computations in \RooFit is the repeated evaluation of the computation graph such as the one 
shown in \cref{PDFTree}. To compute a likelihood, the probability of observing \textit{each} event in the dataset has to be computed.
\RooFit achieves this by loading the values of the observables into the leaves of the computation graph for a \textit{single} event, and evaluating the probability of
the top node. Each node caches its last value, and therefore constant branches of the computation graph (\textit{e.g.} branches that only 
depend on parameters) are only computed once.
Yet, all branches that depend on observables have to be recomputed for each entry in the dataset. In \cref{PDFTree}, for example, the node
"t" in the centre of the graph is the observable, whereas other leaves are parameters. This means that the majority of nodes has to be
recomputed for each entry in the dataset. Evaluating a node involves virtual
function calls, and the number of such calls is proportional both to the number of entries in the dataset and
to the number of (non-constant) nodes in the graph, $N_\mathrm{Data} \cdot N_\mathrm{Nodes}$. For a small graph of 10 nodes and one million
events, this already amounts to considerably more than 10 million function calls because additional calls for the normalisation of PDFs and
for invalidating (node-local) caches are necessary.

Furthermore, loading single values into the nodes of the computation graph is hostile to CPU caches. There is a high likelihood that
when a node is being revisited to load the next entry, data have been thrashed from the highest-level cache(s). This means that \RooFit runs inefficient on
modern CPUs because of poor data locality and inefficient memory access patterns.

\subsection{Batch Computations for Higher Data Locality}
\begin{figure}[t]
\begin{center}
\includegraphics[width=1.\linewidth]{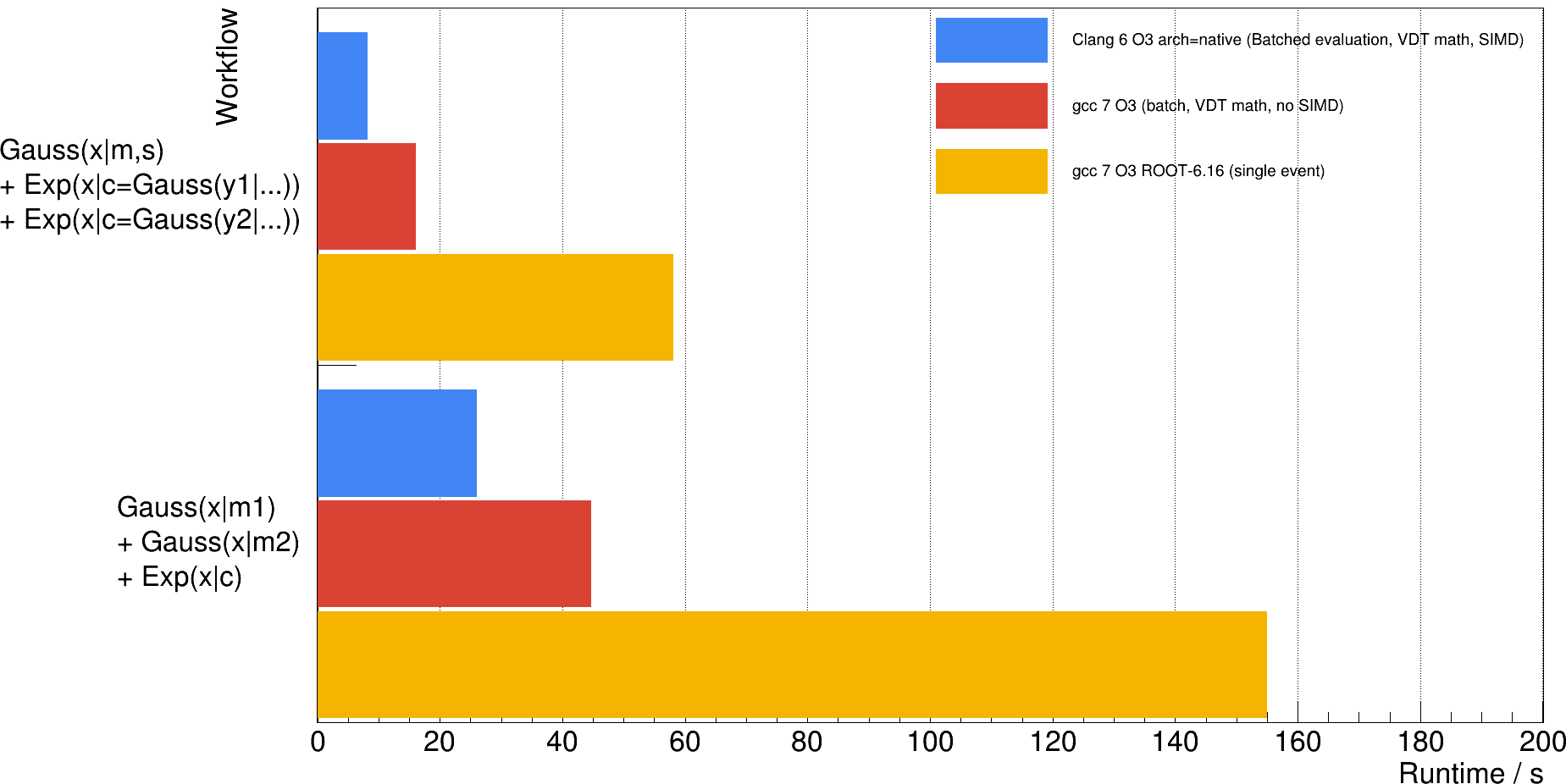}
\end{center}
\caption{\label{VecSpeed}Run time for fitting different likelihood models to datasets of two million events, Intel i7-4790. \textbf{Top}: Using \texttt{AVX2} SIMD instructions and batch data processing increases the speed by $7{\times}$. \textbf{Middle}: Batch data processing leads to a speed up of $3.5{\times}.$ \textbf{Bottom}: Current \RooFit.}
\end{figure}

To demonstrate that a likelihood evaluation can be sped up by loading batches of data, a preliminary interface using \texttt{std::span} was
implemented for a few selected PDFs (Gaussian, Poisson and exponential distribution, summation of PDFs).
Instead of computing only one probability per node, all probabilities for all entries in the data set were
computed in a \textit{single} function call for each node. Since such computations operate on array-like structures, data locality, caching and prefetching improve.
\Cref{VecSpeed} shows that run times for fitting simple models decrease by
a factor $2$ to $3.5$. The speed up is expected to be even larger for larger models.

\subsection{SIMD Computations}
When computations on array-like structures are performed, single-instruction-multiple-data computations (SIMD) can be used. The automatic
vectorisation optimisation of the clang compiler was used to vectorise computations for an \texttt{AVX2} architecture for the Gaussian and exponential distributions, as well as the
addition of PDFs and the normalisation of the three. This requires auto-vectorisable mathematical functions, which need to be inlinable, and
not have any data dependencies between elements of the underlying arrays. Such functions are provided by the
\texttt{VDT}~\cite{VDT} package, of which the logarithm and exponential function were used. Further, computations were rewritten such
that there are no data dependencies, that they can be executed entirely in registers,
and that they use only limited branches, no (non-inlinable) functions and only simple reductions.

Auto vectorisation for these selected PDFs with \texttt{AVX2} instructions  increased the speed up to $6{\times}$ to $7{\times}$.

\section{Summary}
The improvements discussed in \cref{sec:STLCollections,sec:HistFactory} will be released in \texttt{\ROOT 6.18},
and the work on batched and vectorised computations will continue. The batch interface will be refined to work with any PDF, and a fall-back
implementation for PDFs that have not been modified will be provided. Auto-vectorisable batch computations will be implemented for a growing number of PDFs to increase the
single-thread performance of \RooFit by several factors without requiring code changes on the user side.
In conjunction with work an parallelising computations~\cite{Patrick}, a speed up by more than an order of magnitude can be expected.

\vfill

\section*{References}
\bibliography{ACAT19}

\end{document}